\documentclass[12pt]{iopart}

\usepackage{iopams}
\usepackage{color}
\usepackage{multirow}
\usepackage{graphicx}
\usepackage{url}
\usepackage{hyperref}
\hypersetup{backref=true,
 pdfnewwindow=true, colorlinks=true,
 linkcolor=blue, anchorcolor=blue,
 citecolor=blue, filecolor=blue,
 menucolor=blue, urlcolor=blue}

\usepackage{cleveref}

\begin{document}

\title{Terahertz magneto-optical response in ferromagnetic Fe-Co-Al alloys}

\author{Ming Lei}
\address{Chemical and Environmental Engineering, University of California Riverside, California 92521, USA}
\ead{mlei012@ucr.edu}
\author{Sinisa Coh}
\address{Materials Science and Mechanical Engineering, University of California Riverside, California 92521, USA}

\date{\today}

\begin{abstract}
We study the magneto-optical properties of Fe-Co-Al ordered alloys in the terahertz range of frequencies. Using the standard Kubo-based approach to compute intrinsic part of the $\sigma_{xy}(\omega)$ we find a strong dependence of $\sigma_{xy}$ on $\omega$ in the terahertz range. For example, we find that below 10~THz Co$_3$Al has nearly constant $\sigma_{xy}$ and that above 10~THz it is reduced by about 50 times.  Furthermore, we find a strong dependence of $\sigma_{xy}$ on the chemical composition.  For example, we find that the addition of Al to Fe changes the sign of $\sigma_{xy}$, while the addition of Co to Fe leads to a nonmonotonic dependence of $\sigma_{xy}$ on Co concentration.
\end{abstract}

\submitto{\JPCM}
\maketitle

\section{Introduction}

A magnet with a cubic crystal structure and magnetization pointing along the z-axis has a non-zero off-diagonal component of the optical conductivity, $\sigma_{xy} (\omega)$. This contribution to the conductivity occurs because magnetic order in such a system breaks time-reversal symmetry.  The breaking of the time-reversal symmetry is propagated to the electronic degrees of freedom by the spin-orbit interaction.  Therefore, the magnitude of $\sigma_{xy} (\omega)$  is dictated by the spin-orbit interaction strength.\cite{RevModPhys.82.1539} Presence of off-diagonal conductivity $\sigma_{xy}$, and thus in-directly the presence of magnetic order, can be detected optically by comparing the polarization of light incident to, and reflected from, the surface of a magnet. This occurs for example in the so-called magneto-optical Kerr effect, MOKE.\cite{14786447708639245}  Opposite is also true.  If one changes the magnetization direction along the z-axis, $$M_{z} \rightarrow - M_{z},$$ this will result in, $$\sigma_{xy} \rightarrow - \sigma_{xy}.$$ Therefore, one can use the direction of magnetization to control the way in which light reflects from a surface of a magnet.  This effect has been used in memory storage devices\cite{doi:10.1063/1.334915, 1457937} and in the creation of tunable photonic materials.\cite{Atmatzakis:14,0c02440}

At zero frequency, the static conductivity $\sigma_{xy}(\omega=0)$ produces the anomalous Hall effect (AHE).  In materials with very few impurities, so that the diagonal conductivity $\sigma_{xx}$ is above $10^6$~$(\Omega ~ \textrm{cm})^{-1}$, the dominant contribution to AHE originates from the scattering of electrons from the impurities.   On the other hand, somewhat paradoxically, in materials with a moderate amount of impurities, with diagonal conductivity around $10^4$--$10^6$~$(\Omega~\textrm{cm})^{-1}$, the dominant contribution to AHE is intrinsic, independent of the number of impurities.\cite{RevModPhys.82.1539} The intrinsic contribution to AHE is given by the integral of the Berry curvature over the occupied states.\cite{PhysRevLett.93.206602}  However, in the dynamic case, the $\sigma_{xy}(\omega \neq 0)$ can no longer be written as a sum over the Berry curvature, instead it needs to be computed from the Kubo-like sum over empty states, as in Refs.~\cite{PhysRevB.9.4897, uspenskii1989electron,PhysRevB.45.10924,PhysRevB.51.12633}.  Nevertheless, we expect that for small enough $\omega$ the intrinsic contribution still dominates $\sigma_{xy}(\omega)$, which is consistent with findings in Ref.~\cite{matsuda2020room}.

There are many calculations of static intrinsic AHC in the literature.  For example, Yao\cite{PhysRevLett.92.037204} calculated the AHC of ferromagnetic bcc Fe, and the calculated $\sigma_{xy}$ at zero frequency is 751~$(\Omega ~ {\rm cm})^{-1}$. Wang\cite{PhysRevB.76.195109} found that at zero frequency, AHC ($\sigma_{xy}$) in bcc Fe, fcc Ni, and hcp Co is 753, $-$2203, and 477~$(\Omega ~ {\rm cm})^{-1}$, respectively.  Bianco\cite{PhysRevB.90.125153} calculated the AHC in Fe$_{3}$Co, and the $\sigma_{xy}$ value at zero frequency is 452~$(\Omega ~ {\rm cm})^{-1}$. Huang\cite{PhysRevB.91.134409} calculated the AHC of the Heusler compound such as Co$_2$FeAl and found it to be 39~$(\Omega ~ {\rm cm})^{-1}$.

While $\sigma_{xy}$ originates from magnetism, the proportionality coefficient between $\sigma_{xy}$ and the magnetic moment is difficult to predict without performing an explicit first-principles calculation.  As discussed in early work on frequency dependence of $\sigma_{xy}$ from Ref.~\cite{uspenskii1989electron}, as well as in later Refs.~\cite{PhysRevLett.92.037204, PhysRevB.76.195109, PhysRevB.90.125153, PhysRevB.74.195118},  different parts of the reciprocal space can have either positive or negative contribution to $\sigma_{xy}$, as sign will in general depend on position of the Fermi level relative to the subtle spin-orbit induced band gaps in the band-structure.  Therefore, we might expect that, generally $\sigma_{xy}$ will be a very sensitive function of the electronic band structure.  As a consequence, we expect a rich dependence of dynamic $\sigma_{xy} (\omega)$ in the low-frequency range. In particular, we expect that there will be a strong frequency dependence of $\sigma_{xy} (\omega)$ when $\hbar \omega$ is close to the energy of the spin-orbit split bands in the band structure. The energy of the spin-orbit split bands is on the order of tens of meV in ferromagnetic metals such as Fe, Co, or Ni.  These energies lie in the range of frequencies $\sim$10~THz, within the so-called {\it terahertz-gap}: the range of frequencies in the electromagnetic spectrum that are in between the microwave radio frequencies and optical frequencies.

Moreover, we quite generally expect that $\sigma_{xy}$ will be very sensitive on alloying, as even subtle changes in the electronic band structure will change the position of spin-orbit split bands relative to the Fermi level. In this study, we use first-principles techniques to calculate the $\sigma_{xy} (\omega)$ in the THz regime for a specific ternary metallic alloy system, Fe-Co-Al.  We decided to focus on this alloy in particular, as it is known that the addition of Co to Fe leads to strong variations of spin-polarized density of states, as studied, for example, in Ref.~\cite{schoen2016ultra}.  Furthermore, even small addition of Al to Fe--Co can lead to large changes in the measured spin-dependent physical properties, as shown in Refs.~\cite{Childress2011,Valenzuela2011,Kimura2014}.  Our calculations show that alloying is indeed a feasible way to change the THz optical response of ferromagnetic metals in the Fe-Co-Al ternary system. Our results show that alloys like Fe-Co-Al are promising candidates for developing magnetic optical metamaterials, where the manipulation of both magnetic moment orientation and chemical composition serves to regulate their interaction with light.

The early work from Ref.~\cite{uspenskii1989electron} focuses on the frequency dependence of $\sigma_{xy}$ in Fe, Co, and Ni for frequencies $\omega$ above 24~THz (0.1~eV). Most other studies of frequency-dependent $\sigma_{xy}$ are carried out in the optical regime, at even higher energies.  For example, Ref.~\cite{KIM20026} computed $\sigma_{xy}$ for Fe$_{3}$Co and FeCo$_{3}$ in the 120--1300~THz (0.5--5.2~eV) energy range.  Next, Ref.~\cite{kumar2010optical} calculated the optical response of Fe$_{4-x}$Co$_x$ (with $x=1$--$3$) up to the 3100~THz (13~eV) range, while Ref.~\cite{Rhee, ADEBAMBO} reported $\sigma_{xy}$ for FeAl in the 120--1500~THz (0.5--6~eV) range.  Recently, some studies have focused on the off-diagonal optical response in the THz regime. Seifert\cite{202007398} studied the off-diagonal optical response in DyCo$_5$, Co$_{0.32}$Fe$_{0.68}$ and Gd$_{0.27}$Fe$_{0.73}$, and they measured somewhat stronger dependence of $\sigma_{xy}$ on $\omega$ in the range of frequencies below 10~THz.
Matsuda\cite{matsuda2020room} studied the off-diagonal optical response in Weyl antiferromagnet Mn$_3$Sn at very low energy (2.4~THz) and they find weak dependence of $\sigma_{xy}$ on frequency.  On the other hand, calculated and measured $\sigma_{xy}$ in the THz regime is very frequency dependent in SrRuO$_3$, as shown in Refs.~\cite{PhysRevB.81.235218, science.1089408, Shimano_2011}. 

We organize the paper as follows. In Sec.~\ref{sec:methods} we show the calculation methods. In Sec.~\ref{sec:results} we present and analyze our results. We conclude in Sec.~\ref{sec:conclusion}. 

\section{Methods}
\label{sec:methods}

We use the Quantum Espresso package\cite{Giannozzi_2009} to calculate the electronic structure of ordered alloys of Fe, Co, and Al. We use the Generalized Gradient Approximation (GGA) of Perdew, Burke, and Ernzerhof (PBE)\cite{PhysRevLett.77.3865} along with the optimized norm-conserving Vanderbilt (ONCV) pseudopotentials which include spin-orbit interaction.\cite{Hamann2013, SCHLIPF201536, PhysRevB.71.115106}  We choose 120~Ha kinetic-energy cutoff for the plane-wave expansion of the valence wave functions. A 16$\times$16$\times$16 Monkhorst-Pack grid and a smearing\cite{PhysRevLett.82.3296} of 0.01~Ry are used to sample the electron's Brillouin zone. We computed $\sigma_{xy} (\omega)$ using the standard Kubo formula,\cite{JPSJ.12.570}
%
\begin{equation}
\sigma_{\alpha \beta} (\omega)=
\frac{ie^2\hbar}{V N_k}
\lim_{\delta \to 0}
\sum_{\bf{k}}\sum_{n m}\frac{f_{m \bf{k}} - f_{n\bf{k}}}{\varepsilon_{m\bf{k}} - \varepsilon_{n\bf{k} }}\frac{\mathinner{\langle \phi_{n\bf{k}}|} v_{\alpha} \mathinner{|\phi_{m\bf{k}}\rangle\mathinner{\langle \phi_{m\bf{k}}|} v_{\beta} \mathinner{|\phi_{n\bf{k}}\rangle}}}{ \varepsilon_{m\bf{k}} - \varepsilon_{n\bf{k}} - \hbar\omega - i \delta/2} \label{eq:Kubo}
\end{equation}
%
that was already used successfully in Refs.~\cite{PhysRevB.9.4897, uspenskii1989electron,PhysRevB.45.10924,PhysRevB.51.12633}. Here, $\alpha$ and $\beta$ denote Cartesian directions. $V$ is the cell volume. Indices $n$ and $m$ denote different electronic bands. $\bf{k}$ represents the wave vector in the Brillouin zone. $N_k$ is the number of k-points, and $f_{n\bf{k}}$ is the Fermi-Dirac distribution function. $\omega$ is the optical frequency.  We use the Wannier interpolation\cite{RevModPhys.84.1419, MOSTOFI2008685} in evaluating Eq.~\ref{eq:Kubo}, as these calculations require a very dense sampling of the Brillouin zone. We tested the convergence of $\sigma_{xy}$ with the choice of the k mesh. We use adaptive k-mesh refinement to accelerate convergence\cite{PhysRevLett.92.037204} by adding a 5$\times$5$\times$5 fine mesh around regions with a large contribution to $\sigma_{xy}(\omega)$. For pure metals (Fe and Co), a 250$\times$250$\times$250 primary k-mesh is enough to achieve the convergence of $\sigma_{xy}$. For calculations with two atoms per unit cell and four atoms per unit cell, a 200$\times$200$\times$200 and 150$\times$150$\times$150 k-mesh, respectively, was enough to achieve convergence.

In this work we mostly stay within the intrinsic limit $\delta \rightarrow 0$ of the $\sigma_{xy}$ for small $\omega$, as given by Eq.~\ref{eq:Kubo}.  For future work we leave the role of finite electron lifetime, random disorder, or temperature effects on $\sigma_{xy} (\omega)$ in the THz regime.  As discussed in Sec.~\ref{sec:conclusion} we attempt to approximately model disorder within the approach of Ref.~\cite{PhysRevB.82.035104}.

To validate the reliability of our calculation approach, we compare our calculation results with previous calculations. Our calculated $\sigma_{xy}$ of Fe and Co at zero frequency are 758 and 471~$(\Omega ~ {\rm cm})^{-1}$, respectively, which agrees very well with the previous calculation result of 753~$(\Omega ~ {\rm cm})^{-1}$ in Fe and 477~$(\Omega ~ {\rm cm})^{-1}$ in Co.\cite{PhysRevB.76.195109} Furthermore, the calculated $\sigma_{xy}$ of FeCo and Fe$_{3}$Co at zero frequency is 226 and 416~$(\Omega ~ {\rm cm})^{-1}$, respectively. The measured $\sigma_{xy}$ of Fe$_{0.68}$Co$_{0.32}$ at zero frequency is about 350~$(\Omega ~ {\rm cm})^{-1}$\cite{202007398}, which lies between the theoretical values of FeCo and Fe$_{3}$Co. Furthermore, we calculated the Kerr angle of Fe in the range of 0--1.2~eV by using our calculated $\sigma_{xy}$ values and experimental $\sigma_{xx}$ values from Ref.~\cite{krinchik1964magneto,Ordal:83}. Our calculated Kerr angle matches very well with the results in Ref.~\cite{krinchik1968magneto}.

In all of our calculations, we assumed that the magnetization points along the $[001]$ direction.  We did not explore what happens to the off-diagonal conductivity when the magnetization points in any other crystallographic direction, such as $[011]$ or $[111]$.  We decided to restrict our calculations to those with the magnetization axis pointing along the $[001]$ direction, as our goal here is to compare $\sigma_{xy}$ only as a function of the chemical composition of the alloy and frequency.  Furthermore, for consistency, in each calculation we chose the same sign of magnetization along the $[001]$ direction.

\section{Results and discussion}
\label{sec:results}

Now we present our results for ordered Fe-Co-Al alloys. We start by discussing the computed lattice constants and crystal structure of these alloys.  Our results are shown in Table~\ref{table:magnetism}.
Most of the ordered alloys we studied order in the bcc-derived structures B$_2$ and D0$_3$.\cite{Popova_2012,IKEDA2002198,okamoto2016supplemental} The B$_2$ structure is a bcc-derived structure with two atoms in the primitive unit cell.  This structure is therefore present in alloys with the ratio of 1--1 of two elements.   On the other hand, D0$_3$ structure contains four atoms in the primitive unit cell, so it is present in ordered binary alloys with the 1--3 ratio of constituent elements, or the 1--1--2 ratio in the case of ternary alloys. Among the ordered alloys we studied, the only ones that are not in the bcc-derived structure are Al, FeAl$_{3}$, CoAl$_{3}$, Co, and
Co$_{3}$Al.  However, three of these (Al, FeAl$_{3}$, and CoAl$_{3}$) are nonmagnetic, regardless of their crystal structure, so their $\sigma_{xy}$ is identically zero.  Therefore, we don't discuss these cases in more detail.  Next, in the case of Co, we explicitly showed that $\sigma_{xy}(\omega)$ is very similar in bcc and hcp structures.  Therefore, for a more consistent comparison with other members of the Fe-Co-Al family of compounds, we will show results for pure Co in its bcc structure.  Finally, we expect that the remaining exception, Co-rich Co$_3$Al, will also have $\sigma_{xy}(\omega)$ that doesn't strongly depend on the structure, so that we are justified in studying Co$_3$Al in the bcc-derived structure (and not in the lowest energy structure, the fcc-derived L1$_2$).

We find that the computed lattice constants are close to the values experimentally measured.  Small deviations, on the order of 1\% are due to the approximations in our exchange-correlation functional, as well as thermal expansion, as the experimental data in Table~\ref{table:magnetism} are taken at room temperature.

Calculated magnetic moments per atom are also given in Tab.~\ref{table:magnetism}.  Most of these ordered alloys are magnetic in our calculations, with the exception of Al-rich compounds, such as FeAl$_{3}$, CoAl$_{3}$ and CoAl. These findings are in agreement with previous studies.\cite{GONZALESORMENO2002573, 14786435808237047}

\begin{table}
\caption{\label{table:magnetism}Calculated magnetic moment and lattice constant of ordered Fe-Co-Al alloys. The magnetic moment is on a per atom basis for nominally magnetic Fe and Co atoms.}
\begin{center}
\begin{tabular}{llccc}
\hline
&  & $M$ & $a^{\rm calc}$ &  $a^{\rm exp}$ \\
&  & ($\mu_{\rm B} / {\rm atom}$) & (\AA) & (\AA)\\
\hline
 Fe       & bcc  & $2.26$  & 2.84   & 2.86 \cite{ZHU2020} \\
 Fe$_{3}$Co  & D0$_3$ & $2.36$  & 5.71   & 5.71\cite{KIM20026} \\
 FeCo  & B$_2$ & $2.30$  & 2.85   & 2.85\cite{ELLIS1941} \\
 FeCo$_{3}$  & D0$_3$ & $2.02$  & 5.66   & 5.66\cite{KIM20026} \\
 Co       & {\it bcc}  & $1.80$  & 2.82   & 2.82\cite{PhysRevLett.54.1051}\\
 \multirow{2}{*}{Co}   &
 \multirow{2}{*}{hcp}  &
 \multirow{2}{*}{$1.62$}  &
 $a=2.50$ & $a=2.50$ \\
 & & &
 $c=4.04$ & $c=4.09$ \cite{ZHU2020}\\
 \hline
 Fe$_{3}$Al  & D0$_3$ & $1.55$   & 5.76   & 5.79\cite{POPIEL1989127}\\
 FeAl  & B$_2$ & $0.36$  & 2.88   & 2.91\cite{JORDAN2003507}\\
 FeAl$_{3}$  & {\it D0$_3$} & .        & 5.98   & \\
 \hline
 Co$_{3}$Al  & {\it D0$_3$} & $1.05$  & 5.69   & \\
 CoAl  & B$_2$ & .        & 2.86   & 2.86\cite{Cooper1963} \\
 CoAl$_{3}$  & {\it D0$_3$} & .        & 6.01   & \\
 \hline
 Fe$_2$CoAl  & D0$_3$ & $1.54$  & 5.76   & $5.73$\cite{AHMAD2021168449} \\
 FeCo$_2$Al  & D0$_3$ & $1.25$  & 5.71   & $5.73$\cite{Balke2007} \\
 \hline
\end{tabular}
\end{center}
\end{table}

\subsection{Fe and Co}

\begin{figure}[!t]
\begin{center}
\includegraphics{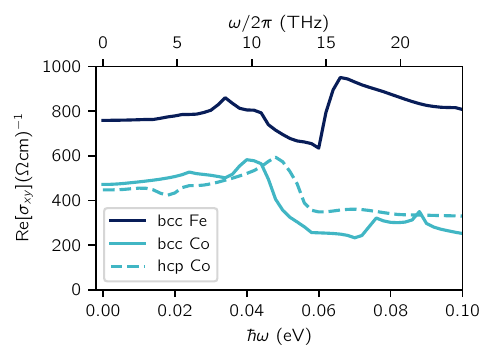}
\end{center}
\caption{\label{fig:co} Calculated real part of $\sigma_{xy}$ for bcc Fe, bcc Co, and hcp Co as a function of $\omega$.}
\end{figure}

Figure~\ref{fig:co} shows calculated $\sigma_{xy} (\omega) $ of pure Fe and Co metal in the range of energies from $\hbar \omega \sim 0$--$0.1$~eV.  This corresponds to the range of frequencies $\omega/(2\pi) \sim 0$--$25$~THz. As expected, we find a strong variation of $\sigma_{xy} (\omega)$ as a function of frequency, in both Fe and Co.  These variations are the strongest around $10$--$15$~THz.  We attribute these modifications to the fact that the characteristic spin-orbit gaps in these metals occur in the same range of energies.  For Fe, we find that the minimal value of $\sigma_{xy}$, in the studied frequency range, is 634~$(\Omega \ {\rm cm})^{-1}$ at $14.5$~THz, while the maximal value is 1.5 times larger, 950~$(\Omega \ {\rm cm})^{-1}$ at a nearby frequency of $16$~THz.  In the case of Co, the minimal value is 233~$(\Omega \ {\rm cm})^{-1}$ at $17$~THz, and maximal value is 2.5 times larger, 582~$(\Omega \ {\rm cm})^{-1}$ at $10$~THz. We compared $\sigma_{xy} (\omega) $ in bcc and hcp structure of Co and we found qualitatively similar responses, as shown in Fig.~\ref{fig:co}.  Therefore, at least in the case of Co, the crystalline structure doesn't have a strong effect on $\sigma_{xy} (\omega)$.

\subsection{Fe-Co alloys}

\begin{figure}[!t]
\begin{center}
\includegraphics{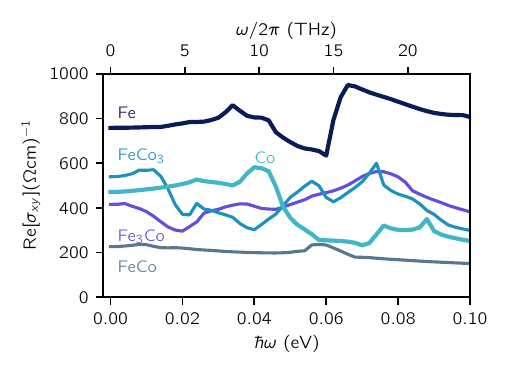}
\end{center}
\caption{\label{fig:adpt-feco} Calculated real part of $\sigma_{xy}$ for bcc Fe, D0$_3$ Fe$_{3}$Co, B$_2$ FeCo, D0$_3$ FeCo$_{3}$ and bcc Co as a function of $\omega$.}
\end{figure}

After analyzing $\sigma_{xy}$ in pure Fe and Co, we now turn to the ordered Fe-Co alloys.  We considered three ordered alloys of Fe and Co, these are Fe$_{3}$Co, FeCo, and FeCo$_{3}$. The ordered crystalline structures of these alloys are bcc-derived D0$_3$, B$_2$ and D0$_3$, respectively.  Our results for Fe-Co alloys are shown in Fig.~\ref{fig:adpt-feco}. Starting from pure Fe, we find that the addition of Co at first significantly reduces the value of $\sigma_{xy}$.  In particular, $\sigma_{xy}$ in Fe$_{3}$Co is on average about 2 times lower than that of pure Fe. Furthermore, the spectral features of alloy  Fe$_{3}$Co are distinct from that of pure Fe. While both Fe and Co have a nearly constant $\sigma_{xy}$ up to 10~THz, we find that Fe$_{3}$Co shows a strong dependence on frequency starting already around 3~THz.  This finding of a large sensitivity of $\sigma_{xy}$ to the chemical composition is in agreement with our expectation that $\sigma_{xy}$ is very sensitive to details of the band structure.  Adding even more Co, as in FeCo, we find that the value of $\sigma_{xy}$ is reduced even further.  In particular, $\sigma_{xy}$ is about 4 times smaller in FeCo than in pure Fe. Moreover, the spectral features of FeCo are surprisingly constant in the entire range of frequencies we studied.  The further addition of Co increases the value of $\sigma_{xy}$.  For example, in the case of FeCo$_{3}$, $\sigma_{xy}$ is similar in magnitude to that of pure Co.  The spectral features of FeCo$_{3}$ are the strongest among the Fe-Co alloys we studied.

The nonmonotonic dependence of $\sigma_{xy}$ on Co concentration is reminiscent of a nonmonotonic dependence of magnetic damping observed in Fe-Co alloys.\cite{schoen2016ultra} Although magnetic damping is not directly related to $\sigma_{xy}$, we expect that both can be related to changes in the nature of the electronic band structure near the Fermi level in Fe-Co alloys.

We note that variation in $\sigma_{xy}$ on Co concentration can't be rationalized with variation of the magnetic moment, as we find that all members of the Fe-Co family of alloys have nearly constant magnetic moment per atom, on the order of $\sim 2 \mu_{\rm B}$ (see Table~\ref{table:magnetism}).  Therefore, even though $\sigma_{xy}$ can be used as a magnetic order signature, one is not necessarily proportional to the other.  As discussed earlier, the role of the magnetic order here is only to break the time-reversal symmetry, while the magnitude and spectral properties of $\sigma_{xy}$ are driven by the spin-orbit interaction of electron bands close to the Fermi level.

Comparing the band structure of all Fe-Co ordered alloys we studied, we find that the overall band structure has not changed much, but only the position of the Fermi level relative to the rigid band structure is increasing with the addition of Co.  Therefore, we attribute the changes in the $\sigma_{xy}$ of Fe-Co alloys to changes in the relative position of the Fermi level, and not to changes in the band structure itself.

\subsection{Fe-Al alloys}

\begin{figure}[!t]
\begin{center}
\includegraphics{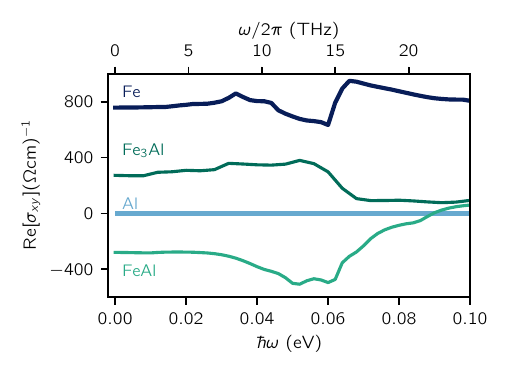}
\end{center}
\caption{\label{fig:adpt-feal} Calculated real part of $\sigma_{xy}$ for bcc Fe, D0$_3$ Fe$_{3}$Al, B$_2$ Fe$_{0.50}$Al$_{0.50}$ and bcc Al as a function of $\omega$.}
\end{figure}

Next, we discuss the calculated $\sigma_{xy}$ in Fe-Al alloys.  Our results are shown in Fig.~\ref{fig:adpt-feal}. Since Al is non-magnetic, the $\sigma_{xy}$ response is, by symmetry, zero at all frequencies. When Al is introduced, the $\sigma_{xy}$ response of Fe$_{3}$Al lies between that of pure Fe and Al, as expected. However,  unexpectedly, when the Al concentration reaches 50\%, $\sigma_{xy}$ response of FeAl changes sign in the entire frequency range from 0 to 20~THz.  We ensured that, in all of these cases, the Fe and Fe-Al alloys have magnetization pointing in the same direction. A similar result is observed in Co/Pd multi-layers in Ref.~\cite{multilayers}, where the sign of $\sigma_{xy}$ changes depending on the relative concentration of Co to Pd.

\subsection{Co-Al alloys}

\begin{figure}[!t]
\begin{center}
\includegraphics{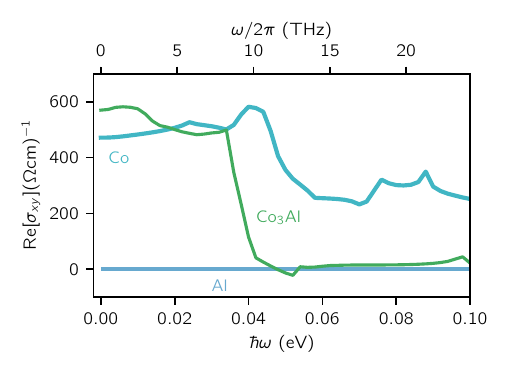}
\end{center}
\caption{\label{fig:adpt-coal} Calculated real part of $\sigma_{xy}$ for bcc Co, D0$_3$ Co$_{3}$Al and bcc Al as a function of  $\omega$.}
\end{figure}

The next binary alloys we discuss are Co-Al alloys.  Results for these alloys are shown in Fig.~\ref{fig:adpt-coal}. Our calculations find that CoAl and CoAl$_{3}$ are not magnetic while the Co-rich compound Co$_{3}$Al is magnetic. At low frequencies, below 10~THz, we find that $\sigma_{xy}$ for Co$_{3}$Al is quite large, and somewhat constant, with the value of $\sim 500$~$(\Omega ~ {\rm cm})^{-1}$.  However, above 10~THz, $\sigma_{xy}$ is reduced 50-fold to only $\sim 10$~$(\Omega ~ {\rm cm})^{-1}$.

\subsection{Fe-Co-Al ternary alloys}

\begin{figure}[h]
\begin{center}
\centerline{\includegraphics{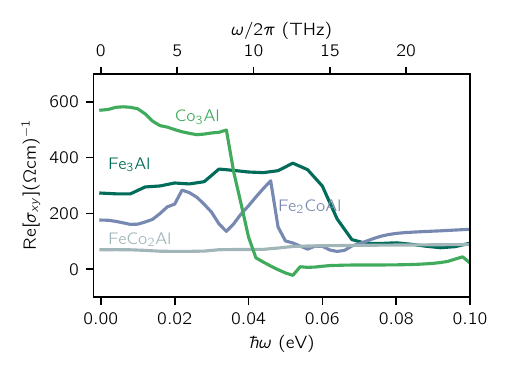}}
\end{center}
\caption{\label{fig:ternary-adpt} Calculated real part of $\sigma_{xy}$ for D0$_3$ Fe$_{3}$Al, D0$_3$ Fe$_2$CoAl, D0$_3$ FeCo$_2$Al and D0$_3$ Co$_{3}$Al as a function of  $\omega$.}
\end{figure}

So far, we discussed the $\sigma_{xy}$ response of binary alloys. In what follows, we consider several Fe-Co-Al ternary alloys.  We kept the Al concentration at 25\%, and varied the relative concentration of Fe and Co. The results for these alloys are shown in Fig.~\ref{fig:ternary-adpt}. The behavior of Fe$_2$CoAl is qualitatively similar to that of Fe$_3$Al, so the replacement of Fe with Co did not change qualitatively $\sigma_{xy} (\omega)$.  Quantitatively, we find that $\sigma_{xy}$ is approximately $\sim 1.5$ times lower in Fe$_2$CoAl relative to Fe$_{3}$Al.  However, with an even higher concentration of Co, as in FeCo$_2$Al, we find $\sigma_{xy} (\omega)$ which is qualitatively and quantitatively different from other members of the Fe-Co-Al family of compounds.  In particular, we find an unusually small $\sigma_{xy}$ in FeCo$_2$Al, around 80~$(\Omega ~ {\rm cm})^{-1}$, that is almost insensitive to the frequency $\omega$ in the entire range between 0 and 25~THz.

\subsection{Optical conductivity and magnetic moment}

\begin{figure}
\begin{center}
\centerline{\includegraphics{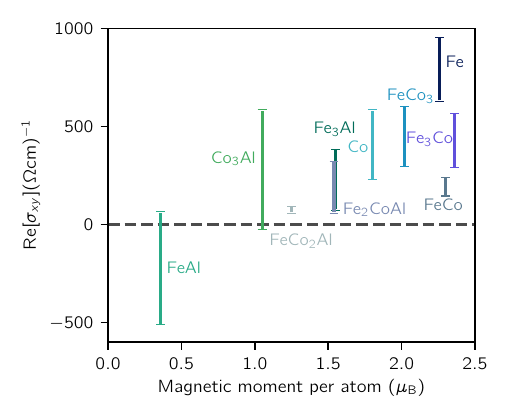}}
\end{center}
\caption{\label{fig:relation} The summary of $\sigma_{xy}$ in the range of 0.0--0.1~eV and magnetic moment per atom of each metal and alloy. The bar shows the maximum and minimum of the $\sigma_{xy}$ in the range of 0.0--0.1~eV.}
\end{figure}

As we discussed earlier, $\sigma_{xy}$ is often taken as an optical signature of magnetic order.  Therefore, it is natural to ask whether a material with a large magnetization will also have a large $\sigma_{xy}$. Figure~\ref{fig:relation} shows the relationship between $\sigma_{xy}$ and the magnetic moment per atom for all of the compounds we studied.  For each compound we show on the vertical scale of Fig.~\ref{fig:relation} the range of maximal and minimal values of the calculated $\sigma_{xy}$ in the range of frequencies from 0 to 25~THz. As can be seen from the figure, we find that in the case of Fe-Co alloys, the magnetic moments per atom are nearly the same, on the order of $\sim 2$~$\mu_{\rm B}$, but the $\sigma_{xy}$ ranges from 150, all the way to 950~$(\Omega ~ {\rm cm})^{-1}$. The same is true for the other compounds we studied.  The most drastic example is FeAl in which $\sigma_{xy}$ even changes sign relative to that of Fe. Therefore we can conclude that materials with larger magnetic moment do not necessarily have larger $\sigma_{xy}$.  This is not too surprising, as the large magnetization arises from the large difference in population of dominantly spin-up and spin-down bands. However, a large $\sigma_{xy}$ in the low-frequency regime relies on detailed information about spin orbit split bands near the Fermi level.

\subsection{Imaginary part of the off-diagonal component $\sigma_{xy}$}

So far we have focused on the frequency dependence of the real part of $\sigma_{xy}$ in Fe-Co-Al ternary system.  However, there are other changes to the conductivity tensor induced by the magnetic order.  These are the imaginary part of the off-diagonal component $\sigma_{xy}$, as well as the difference between the diagonal components of the conductivity tensor along and perpendicular to the direction of the magnetic moment, $\sigma_{zz} - \sigma_{xx}$.

We first briefly discuss the imaginary part of $\sigma_{xy}$ which contributes to the magneto-optical Kerr effect, MOKE, in addition to the real part of $\sigma_{xy}$. The imaginary part of $\sigma_{xy}$ is shown in the supplement for all the compounds we studied.  As expected from the Kramers-Kronig relationship, the imaginary part of $\sigma_{xy}$ is also strongly dependent on both chemical composition and frequency.  For example, we find that the step-like spectral feature we found for the real part of $\sigma_{xy}$ in Co$_3$Al is accompanied with a single sharp feature in the imaginary part of $\sigma_{xy}$, at nearly the same frequency ($\sim 10$~THz), as one might expect.  Similarly, sharp features in the imaginary part of $\sigma_{xy}$ we find in the case of Co and Fe$_3$Al.  The spectral features of the imaginary part of $\sigma_{xy}$ are much more complex for the other compounds we studied.  This is particularly true for alloys containing Fe magnetic atoms.

\subsection{$\sigma_{zz} - \sigma_{xx}$}

\begin{figure}
\begin{center}
\centerline{\includegraphics{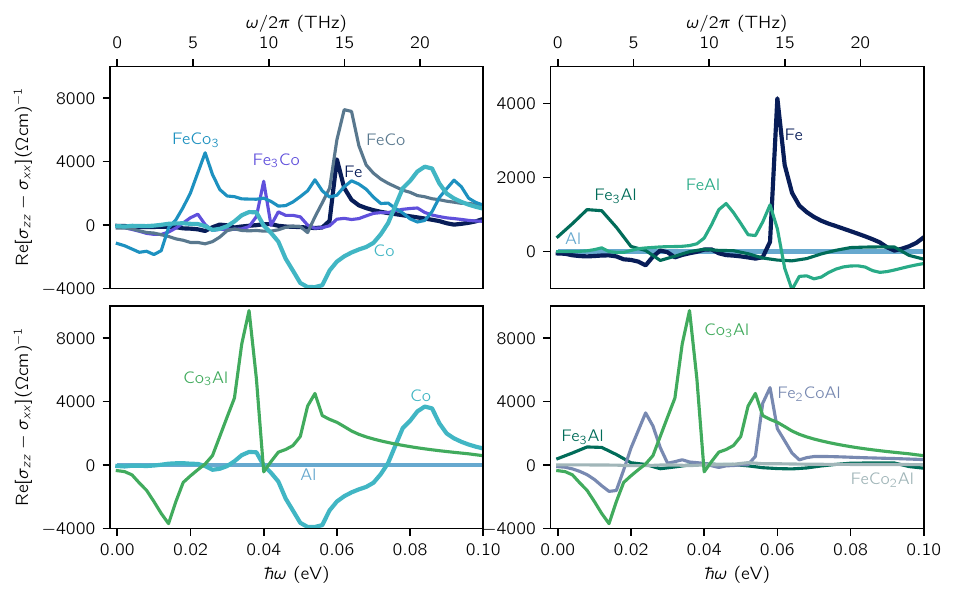}}
\end{center}
\caption{\label{fig:zz-xx}  Calculated real part of $\sigma_{zz}$-$\sigma_{xx}$ for Fe-Co-Al metals and alloys as a function of frequency $\omega$.  
}
\end{figure}

For completeness, we now analyze the remaining component of the conductivity tensor that depends on the presence of the magnetic order.  By symmetry, magnetization along the $z$ axis introduces a spin-orbit driven difference between the $\sigma_{zz}$ (along the magnetization axis) and $\sigma_{xx}=\sigma_{yy}$ (perpendicular to the magnetization axis). While the difference $\sigma_{zz} - \sigma_{xx}$ does not contribute to the magneto-optical Kerr effect (MOKE) it does contribute to the second order change in the birefringence (Voigt effect).  While $\sigma_{xy}$ is zero without magnetic order, the diagonal components $\sigma_{xx}=\sigma_{yy}$ and $\sigma_{zz}$ are not. Therefore, here we don't focus on these diagonal components individually, but instead we focus on their difference, $\sigma_{zz}-\sigma_{xx}$.  The calculated values of $\sigma_{zz}-\sigma_{xx}$ for all compounds we studied are shown in Figure~\ref{fig:zz-xx}.  Again, we find strong variations of $\sigma_{zz}-\sigma_{xx}$ both as a function of frequency and as a function of chemical composition.  In the case of the Fe-Co family of compounds, we find that $\sigma_{zz}-\sigma_{xx}$ in Fe and FeCo have a sharp peak around 15~THz, and are nearly zero below 14~THz. Co shows two sharp features, with opposite signs, one at around 9 and another around 20~THz.  The addition of 25\% of Al to Fe strongly changes the value of $\sigma_{zz}-\sigma_{xx}$.  While Fe has a sharp feature around 15~THz, Fe$_3$Al has a sharp feature around 3~THz.  The addition of even more Al, as in FeAl, introduces two sharp features, at around 10 and 15~THz.  The addition of Al to Co also introduces significant qualitative and quantitative changes to $\sigma_{zz}-\sigma_{xx}$.

\section{Discussion and conclusion}
\label{sec:conclusion}

We find a strong dependence of $\sigma_{xy} (\omega)$ in the terahertz frequency range for Fe-Co-Al ordered alloys. For example, in the case of Co$_3$Al we find a nearly 50-fold reduction in $\sigma_{xy} (\omega)$ above 10~THz as compared to $\sigma_{xy} (\omega)$ below 10~THz.  On the other hand, in the case of FeCo$_2$Al we find a nearly constant $\sigma_{xy} (\omega)$ in the entire range from 0 to 25~THz. Furthermore, we also find a strong dependence of $\sigma_{xy} (\omega)$ on the chemical composition.  For example, the addition of Al to Fe can change the sign of $\sigma_{xy} (\omega)$, so that $\sigma_{xy} (\omega)$ is positive in Fe and Fe$_3$Al but negative in the case of FeAl.  Similarly, the addition of Co to Fe produces a nonmonotonic dependence of $\sigma_{xy} (\omega)$ on Co concentration.  As an example, $\sigma_{xy} (\omega)$ in FeCo is about 4 times smaller than in Fe and 3 times smaller than that in Co.  

We attribute both of these strong variations, with frequency and composition, to the changes in the electronic structure induced by the presence of the spin-orbit interaction.  Detailed, k-space resolved, analysis of $\sigma_{xy}(\omega)$ is rather involved, even in the case of pure Fe, at $\omega = 0$. For example, Ref.~\cite{PhysRevB.74.195118} found that the $\sigma_{xy} (\omega=0)$ of pure Fe is contributed both by small regions of k-space, close to avoided crossings, which contribute to the AHE with high intensity, and equally important, background contributions with smaller intensity, but present at nearly all k-points. The analysis is additionally complicated in our work, as we study binary and ternary alloys, and we study $\omega > 0$ which can't be written as a sum over the occupied states. Therefore, we leave a more detailed analysis of the origin of these strong variations of $\sigma_{xy} (\omega)$ to future studies. 

Instead, here we focus on a different analysis, inspired by the findings of Ref.~\cite{PhysRevB.74.195118}.  Our analysis is based on the joint density of states (JDOS).  Such an analysis sums over all k points and can therefore be equally well applied to pure metals as well as to alloys, regardless of the number of atoms in the primitive unit cell.  Furthermore, this analysis is particularly well suited for analysis of $\sigma_{xy}$ at nonzero frequencies $\omega$, since JDOS counts how many states are available for transition at a particular frequency $\omega$.

\begin{figure}
\begin{center}
\centerline{\includegraphics{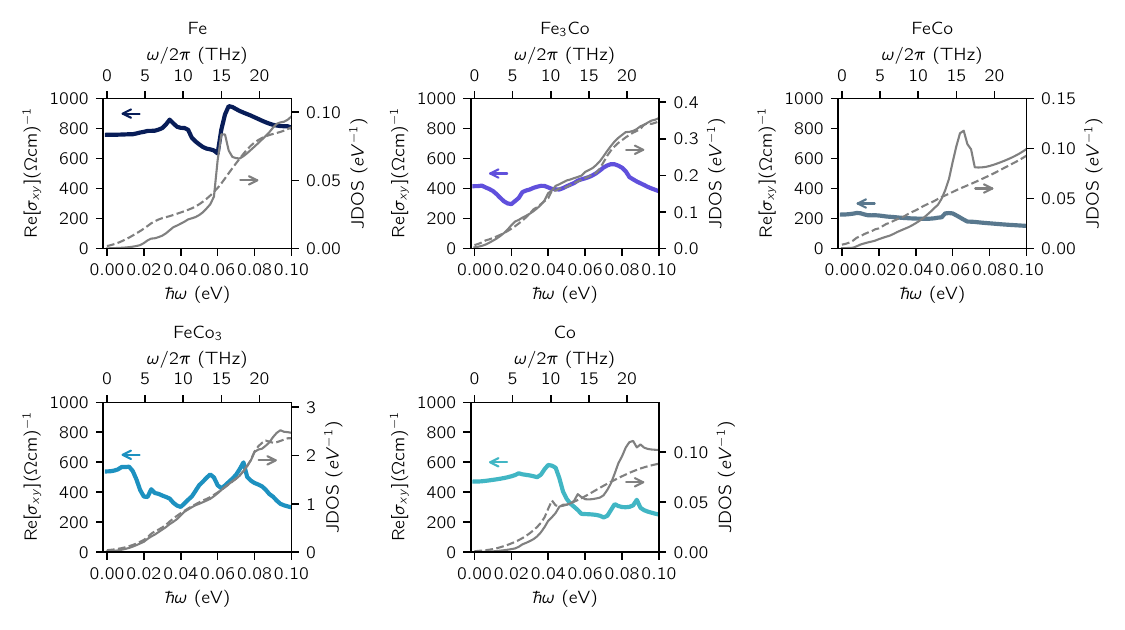}}
\end{center}
\caption{\label{fig:jdos-feco} Calculated joint density of states of Fe-Co  alloys as a function of frequency $\omega$. The grey lines show the density of states with the spin-orbit coupling (solid gray line) and without the spin-orbit coupling (dashed gray line).  Spectral features present in the solid gray line, but not in the dashed gray line, can therefore be assigned to the presence of the spin-orbit interaction.  The colored lines show the real part of $\sigma_{xy}$ in the same frequency range as the joint density of states.}
\end{figure}

\begin{figure}
\begin{center}
\centerline{\includegraphics{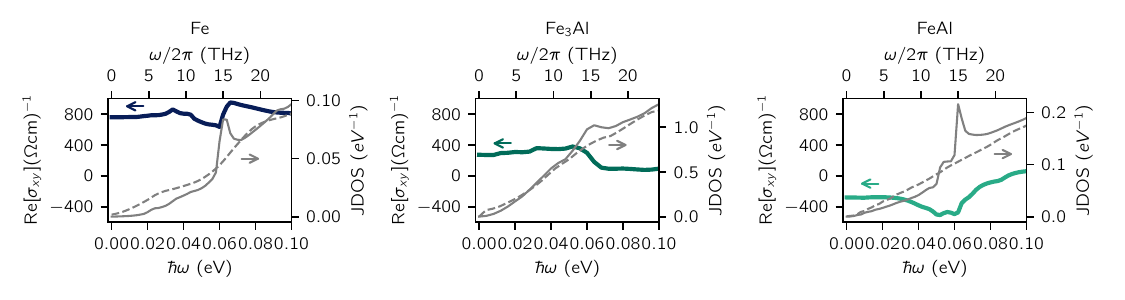}}
\end{center}
\caption{\label{fig:jdos-feal} Calculated joint density of states of Fe-Al as a function of frequency $\omega$. All conventions are the same as in Figure~\ref{fig:jdos-feco}.}
\end{figure}

\begin{figure}
\begin{center}
\centerline{\includegraphics{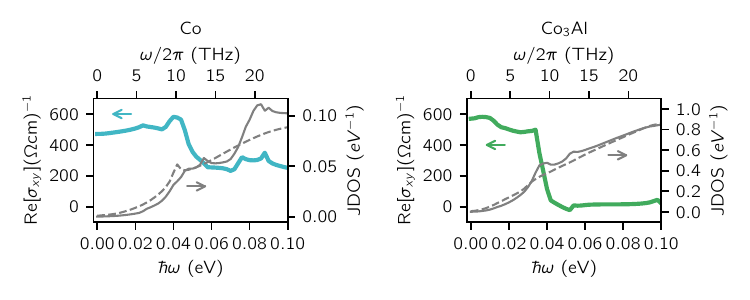}}
\end{center}
\caption{\label{fig:jdos-coal}  Calculated joint density of states of Co-Al as a function of frequency $\omega$. All conventions are the same as in Figure~\ref{fig:jdos-feco}.}
\end{figure}

\begin{figure}
\begin{center}
\centerline{\includegraphics{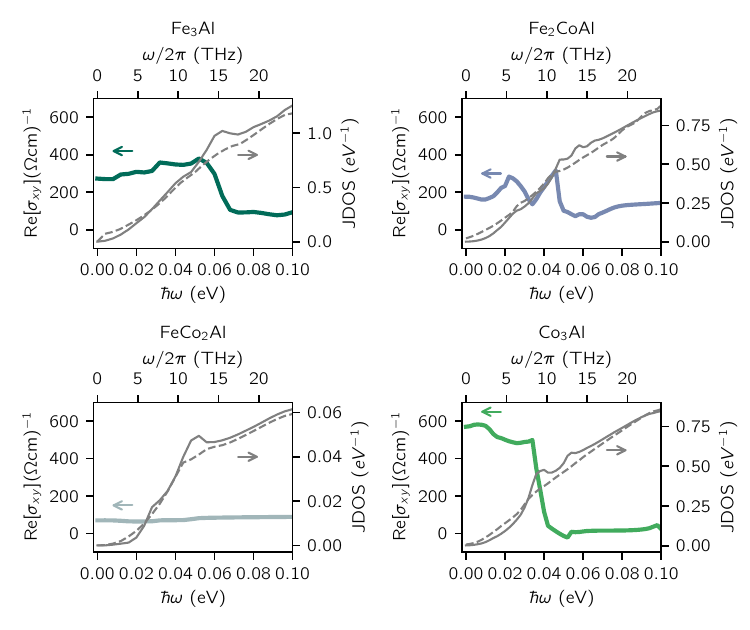}}
\end{center}
\caption{\label{fig:jdos-ternary} Calculated joint density of states of Fe-Co-Al alloys as a function of frequency $\omega$. All conventions are the same as in Figure~\ref{fig:jdos-feco}.}
\end{figure}

We start our analysis by first computing JDOS without including spin-orbit interaction in the calculation.  Next, we include spin-orbit interaction in the calculation, and compute the JDOS once again.  Comparing the JDOS between those two calculations tells us what is the effect of the spin orbit on the electronic spectrum. Clearly, any changes in the electronic spectrum induced by spin-orbit interaction should correlate with $\sigma_{xy}$, as $\sigma_{xy}$ originates purely from the spin-orbit interaction.  We then expect that the frequencies $\omega$ where the two JDOS differ will reflect in $\sigma_{xy} (\omega)$, as this is the part of the spectrum where the spin-orbit interaction has redistributed the weight of electronic states. We report the joint density of states with and without inclusion of the spin-orbit interaction in Figures~\ref{fig:jdos-feco}, \ref{fig:jdos-feal}, \ref{fig:jdos-coal}, \ref{fig:jdos-ternary}. Dashed gray lines show JDOS without spin-orbit interaction, whereas solid gray lines show JDOS with spin-orbit interaction.  As can be seen in the figures, we find that whenever $\sigma_{xy}(\omega)$ experiences a large change as a function of $\omega$, there is a corresponding spike in the JDOS due to the inclusion of the spin-orbit interaction.  Therefore, the gaps in the electron spectrum induced by the spin-orbit interaction are well correlated to $\sigma_{xy}(\omega)$, as expected.

Our calculations are performed in the limit of infinite electron lifetime.  While this is justified for materials with moderate amount of disorder, with diagonal conductivity around $10^4$--$10^6$~$(\Omega~\textrm{cm})^{-1}$, we leave for future work discussion of role of disorder on $\sigma_{xy} (\omega)$ in the terahertz range for materials that are not within the moderate range of disorder.  We expect that at low enough frequencies the phenomenology of $\sigma_{xy} (\omega)$ will be the same as that of $\sigma_{xy} (\omega=0)$, so that with a moderate amount of disorder the dominant contribution to $\sigma_{xy} (\omega)$ is intrinsic, but with less disorder scattering from impurities start to dominate.\cite{RevModPhys.82.1539}  We show in supplement $\sigma_{xy} (\omega)$ with approximately incorporated effect of the finite carrier lifetime.  While the finite carrier lifetime approximation washes out some of the spectral features in $\sigma_{xy} (\omega)$, we still find that many qualitative characteristics remain, such as the change in sign of $\sigma_{xy}$ near $20$~THz, or the sharp decline in $\sigma_{xy}$ in Co$_3$Al around 10~THz, or non-monotonic dependence of $\sigma_{xy}$ of Co concentration in Fe-Co alloys.

Our findings indicate that alloys such as Fe-Co-Al would be of interest in the creation of magnetic optical metamaterials in which the direction of magnetic moment and chemical composition are used to control its interaction with light.

\ack{This work was supported by grant NSF DMR-1848074. Computations were performed using the HPCC computer cluster at UCR. We acknowledge discussion with Igor Barsukov and Richard B. Wilson.}

\newcommand{\newblock}{}

\bibliographystyle{unsrt}

\bibliography{pap}


\end{document}